\documentclass{aajos}

\usepackage{epsfig,graphicx}
\usepackage{natbib}
\bibpunct{(}{)}{;}{a}{}{,}    
\usepackage{amsmath} 
\usepackage{mathrsfs}
\usepackage[varg]{txfonts}

\begin{document}

\title{Dark energy as an emergent phenomenon}
\titlerunning{Dark energy as an emergent phenomenon}
\author{J.O. Stenflo}

\institute{Institute for Particle Physics and Astrophysics, ETH
  Zurich, CH-8093 Zurich \\ \email{stenflo@astro.phys.ethz.ch} \and 
  Istituto Ricerche Solari Locarno (IRSOL), Via Patocchi, CH-6605 Locarno-Monti, Switzerland}


\abstract{The cosmological constant,
which was introduced by Einstein a century ago to allow for a static
universe, experienced a revival two decades ago under the label dark
energy as a parameter to
model the observed accelerated expansion of the universe. Its 
physical nature has however remained enigmatic. Here we use the 
Einstein equations without cosmological constant to 
show that the origin of the accelerated expansion is not in the
equations but has to do with the 
boundary conditions related to the causal horizon, which exists
because the age of the universe is finite. Via transformations to
conformal coordinates and Euclidian spacetime we find a resonance
condition that uniquely determines the dimensionless parameter 
$\Omega_\Lambda$ that governs the observed cosmic acceleration: $\Omega_\Lambda
=\textstyle{2\over 3}\,(\pi\, t_H/t_c)^2$, where $t_H$ is the Hubble
time and $t_c$ is the conformal age of the universe. This explanation
leads to a somewhat modified cosmology, in which the expansion rate of
the early universe is 2.1 times faster than in the standard
model. We show that Big
Bang nucleosynthesis calculations with the faster expansion rate
requires the mean baryon density to be raised to the level of the
total matter density 
to agree with the observed deuterium abundance. This appears to
eliminate the need to invoke the existence of some yet to be discovered exotic
particles to explain dark matter, since all of it may be baryonic
while still remaining consistent with the observed abundances of the light elements. 
}


\maketitle


\section{Introduction}\label{sec:intro}
According to standard cosmology the universe is 13.8 billion years old
and spatially flat, but only 4.9\,\%\ of the critical mean energy
density that is required for flatness is in the form of ordinary matter,
baryons. The rest of the energy density resides in two enigmatic
components, 25.9\,\%\ as dark matter and 69.1\,\%\ as dark energy
\citep{stenflo-planck2016}. While the discovery of dark matter goes
back to \citet{stenflo-zwicky1937} and remains a mystery eight decades
later, dark energy was introduced two decades ago to
account for the accelerated expansion of the universe that was revealed by
the use of supernovae type Ia as standard candles
\citep{stenflo-riessetal1998,stenflo-perlmutteretal1999}. The physical
nature of dark energy also remains a complete mystery and is being
treated as a fitting parameter in the standard cosmological models, in
the form of a cosmological constant. 

\citet{stenflo-einstein17} introduced the cosmological constant to
allow for a static universe but considered it a 
blunder after observations revealed that the universe was not static
but expanding. All later observations were compatible with a zero
cosmological constant $\Lambda$ until the accelerated expansion was
discovered. $\Lambda$ represents a property of the
spacetime metric in the form of a vacuum energy that
one would expect in quantum field theories (QFT). It can therefore
be expected to represent the vacuum energy in a theory of quantum
gravity. The problem is that estimates based on straightforward dimensional
analysis lead to a predicted value for $\Lambda$ that is about 122 orders of
magnitude larger than the presently observed value (cf. 
Eq.~(\ref{eq:rhoscaling}) below). This monumental discrepancy illustrates
how far we are from an understanding of the gravitational role of the
quantum vacuum. It has led to the belief that a resolution of the dark
energy enigma will require a theory of quantum gravity
\citep[cf.][]{stenflo-binetruy2013}. 

If the vacuum energy were as large as expected from QFT, and if this
energy were a source of gravity, then the universe would collapse on a
miniscule time scale without the opportunity to grow large to allow
a sufficient time scale for biological evolution. Long before the
accelerated expansion was discovered \citet{stenflo-weinberg1987}
invoked the anthropic principle to set tight upper limits on the
magnitude of $\Lambda$, which would be compatible with our existence
as observers. While such an argument effectively restricted the
allowed range, it opened the door to the possibility of parallel
universes with other values of $\Lambda$, most of which (an infinity
of them) with values incompatible with the existence of life. String
theory allows a ``landscape'' of $10^{500}$ possible universes,
and as no theoretical procedure to choose among them has yet been
found, the anthropic argument has again been invoked and presented as
if it would offer an explanation for the kind of universe that we live
in \citep[cf.][]{stenflo-ellis2009}. 

Before the discovery of accelerated expansion it seemed that the mean
energy density of the universe was much smaller than the critical
density required for spatial flatness, even when accounting for all
the invisible dark matter. This would be incompatible with inflation,
which predicts that the present universe must have nearly zero
curvature. The discovered dark energy filled the gap. All 
forms of energy, baryonic, non-baryonic, and vacuum energy, now add up
to the critical density, thus restoring flatness. 

One major conceptual problem with a cosmological constant is that it
leads to a cosmology that is in gross violation of the Copernican
principle, which states that we are not privileged observers in the
universe. In the cosmological context it is often referred to as the
cosmic coincidence problem. When the universe expands as described by a
scale factor $a(t)$, $\Lambda$ stays constant while the densities of
matter and radiation vary as $a^{-3}$ and $a^{-4}$, respectively. This
implies that the radiation energy density in the Planck era dominated
over the vacuum energy density by 122 orders of magnitude, while in
the future it is the vacuum energy density that will dominate by an
increasing number of orders of magnitude. We happen to live in an
epoch when the vacuum and matter energy densities are of the same
order. This extraordinary coincidence has not been given any
explanation other than again referring to the anthropic principle. 

Let us now turn from dark energy to dark matter. Evidence for its
existence comes from a variety of 
observations, the most important ones being the rotation curves of
galaxies, the velocity dispersion of galaxies in clusters, and
gravitational lensing. While explanations in terms of modified gravity
have been tried, the general consensus is now that dark matter really
exists in the form of particles. Particularly convincing evidence for
this comes from observations of a collision between two galaxy
clusters, the Bullet Cluster, which shows that the visible component
is significantly displaced by the collision relative to the invisible
(but gravitating) dark matter component, something that has no
explanation in terms of modified gravity
\citep[cf.][]{stenflo-bullet2004,stenflo-bullet2006}. 

It is clear that dark matter must consist of cold (non-relativistic), dark
collisionless matter, but this does not exclude that it can be
baryonic. The main reason why it is widely believed to be non-baryonic
comes from the constraints of BBN (Big Bang nucleosynthesis)
calculations. Only about 5\,\%\ of the critical density of the
universe can be in baryonic form if the BBN predictions are to agree
with the observed abundances of the light elements. Since the total
matter fraction is about 30\,\%, the implication is that there is
about 5 times more non-baryonic than baryonic matter. Often the
non-baryonic matter is referred to as WIMPs, weakly interacting
massive particles. 

Nobody knows what kind of particles WIMPs are made of, although major
search efforts have been carried out for several decades. The searches
are made with large underground detectors in order to filter out
spurious signals from cosmic-ray particles. Attempts to produce
hypothetical dark matter particles by colliders like LHC at CERN have
also been unsuccessful so far. As time goes on without anything else
than null results, the credibility
of the belief that most of the dark matter is in an
exotic non-baryonic form suffers. However, as good
alternative explanations of dark matter are unavailable, the search
continues unabated. 

The aim of the present paper is to show that the unaltered Einstein
equations without a cosmological constant lead to an accelerated
expansion of the universe of the observed magnitude, as a consequence of the
boundary conditions that must be enforced to preserve consistency. The
observable universe is bounded by being enclosed inside a causal
horizon, which exists because the age of the universe is finite. 
The causal boundary constraint leads to a resonance condition. Because this
condition is always tied to the size of the causal horizon, the
coincidence problem disappears. The main resulting modification of the
cosmological evolution is an expansion rate in the early universe that
is 2.1 times faster than in the standard model. With this faster rate
the BBN predictions appear to give agreement with the observed abundances of the
light elements only if the baryonic mean density is increased from
5\,\%\ to values of order 30\,\%, which would eliminate the need to
invoke the existence of non-baryonic matter or WIMPs to 
account for these abundances. 

In Sect.~\ref{sec:gravpot} we first explore the properties of the gravitational
potential in the presence of a cosmological constant. The Newtonian
potential gets changed into a Helmholtz 
potential that represents the solution of a wave equation. The spatial
scale of the wave is related to the radius of the causal horizon. In
Sect.~\ref{sec:origin} 
we then remove the cosmological constant from the
equations and determine the resonance condition that is induced by the
presence of the causal horizon. This requires the use of conformal
coordinates, with which the expansion factor $a(t)$ is transformed away, as
well as the transformation to Euclidian spacetime. The resonance
condition uniquely determines the value of $\Lambda$ without the use
of any free parameters. In Sect.~\ref{sec:darkmatter} we use a
simplified BBN treatment of deuterium production in the early universe
to show that with our non-standard enhanced expansion rate the baryon
density needs to be enhanced to the level of the total matter density to preserve
agreement with the observed deuterium abundance. We finally summarize
the conclusions in  Sect.~\ref{sec:concl}.


\section{Gravitational potential in the presence of a cosmological
  constant}\label{sec:gravpot}  
The present paper aims at explaining dark energy as an emergent
phenomenon that is not explicitly present in the underlying equations 
for the metric and the gravitational field. Before addressing the question
of its origin, let us here start by taking a look at the roles played by the cosmological
$\Lambda$ term when it is inserted ad hoc in the Einstein equations in
the standard way. This term has two main effects. While its role of a
vacuum energy density (dark energy) acting as repulsive gravity is
well known, its second role as a kind of vacuum polarization
representing a feedback of the vacuum energy on the
gravitational 
interaction has been largely overlooked. In the present section we
will highlight this feedback and show how it leads to a wave
equation for the interaction. 


\subsection{Poisson equation for the gravitational
  potential}\label{sec:poisson} 
Newtonian gravity faces serious inconsistency problems when trying to
deal with an infinite distribution of matter. If one tries 
to apply the shell theorem, the force on a particle depends on
the arbitrary choice of center for the spherical
geometry. As described by \citet{stenflo-ghosh16}, already
\citet{stenflo-laplace1880} tried to deal with this problem by introducing a Yukawa-like
exponential cut-off of the gravitational potential $\Phi$: 
\begin{equation}
\Phi \sim -e^{-r/r_s}/r\,,
\label{eq:scpot}\end{equation}
where $r_s$ is the screening or cut-off distance that defines the
finite range of the gravitational force. 

Such a Yukawa-like potential was also applied for similar reasons by
\citet{stenflo-seeliger1895} and \citet{stenflo-neumann1896}. It results from the
screened Poisson equation  
\begin{equation}
\nabla^2 \Phi -\lambda\,\Phi =4\pi \,G\,\rho\,,
\label{eq:scpoisson}\end{equation}
which served as Einstein's Newtonian starting point when he introduced his
cosmological constant $\lambda$ in his 1917 cosmological paper
\citep{stenflo-einstein17}. 

There has since been considerable confusion whether or not a positive
cosmological constant really leads to a Yukawa-like gravitational
potential. Thus \citet{stenflo-straumann02} points out that Einstein,
Weyl, and Pauli saw the cosmological term 
as a Yukawa term, but he then argues that this interpretation is
incorrect, since the stationary solution for $\Phi$ given by general
relativity for a homogeneous universe is 
\begin{equation}
\nabla^2 \Phi = 4\pi\, G\,(\rho -2\rho_\Lambda)\,,
\label{eq:poissuniv}\end{equation}
where $\rho$ is the mean matter density, while 
\begin{equation}
\rho_\Lambda\equiv {c^2\Lambda\over 8\pi\, G}
\label{eq:rholam}\end{equation}
represents the vacuum energy density, with $\Lambda$ being
the cosmological constant. When the vacuum energy term in 
Eq.~(\ref{eq:poissuniv}) is moved over to the left-hand side, we see that
its sign is opposite to that of the corresponding term in 
Eq.~(\ref{eq:scpoisson}), and this has been taken to imply that the
gravitational potential would not be screened. This conclusion has
been restated in the nice review by \citet{stenflo-raifear17}. 

However, when comparing Eqs.~(\ref{eq:scpot})-(\ref{eq:poissuniv}) with each other, we
notice that there is something profoundly missing. While the vacuum energy term in 
Eq.~(\ref{eq:poissuniv}) is a constant, independent of space and time, 
$\lambda$ in Eq.~(\ref{eq:scpoisson}) is a multiplicative factor for the $r$
dependent potential $\Phi$ in Eq.~(\ref{eq:scpot}). The $\lambda$
term therefore induces a feedback from $\Phi$ to its own spatial
gradients in the screened Poisson equation, and it is this feedback
that is the reason for the screening. Regardless of the sign issue,
this feedback is missing in Eq.~(\ref{eq:poissuniv}).


\subsection{Derivation in general relativity}\label{sec:standder} 
To understand the origin of this feedback, let us here
provide a brief derivation, starting with the standard formulation of
the Einstein field equation with a cosmological constant $\Lambda$ term: 
\begin{equation}
R_{\mu\nu} -{1\over 2} \,g_{\mu\nu} R +\Lambda\, g_{\mu\nu}=-\,{8\pi \,G\over c^4} \,\,T_{\mu\nu}\,.
\label{eq:einstein1}\end{equation}
Our treatment will be based on the convention ($+--\,-$) for the
spacetime signature. 

Before introducing the weak-field approximation it is convenient to
rewrite Eq.~(\ref{eq:einstein1}) in the form 
\begin{equation}
R_{\mu\nu} =-\,{8\pi \,G\over c^4} \,\,S_{\mu\nu}\,,
\label{eq:einstein2}\end{equation}
where
\begin{equation}
S_{\mu\nu} =T_{\mu\nu} -{1\over 2} \,g_{\mu\nu} T -\,{c^4\,\Lambda\,
  g_{\mu\nu}\over 8\pi \,G}\,.
\label{eq:smunu}\end{equation}
This form is readily obtained from Eq.~(\ref{eq:einstein1}) in
the standard way by making contraction with $g^{\mu\nu}$ and using the
circumstance that $g^{\mu\nu}$ contracted with $g_{\mu\nu}$ equals 4. 

Let us write $g_{\mu\nu} \equiv 
\eta_{\mu\nu} + h_{\mu\nu}$, where $\eta_{\mu\nu}$ represents the
invariant Minkowski metric with diagonal elements unity and signs
$+--\,-$, while $h_{\mu\nu}$ is the spacetime dependent part of the
metric. In the weak-field approximation  $\vert h_{\mu\nu}\vert\ll 1$,
and $R_{\mu\nu}\approx
{1\over 2}\,\partial^2 g_{\mu\nu}\,$ if we, as is usually done, adopt
the harmonic gauge. $\partial^2\equiv(1/c^2)\,\partial^2/\partial t^2 -\nabla^2$ is
the 4D d'Alembertian 
operator. Its spatial part is the negative Laplace
operator $-\nabla^2$. We further assume here for simplicity that the
stress-energy tensor $T_{\mu\nu}$ consists of baryonic matter with zero
equation of state (zero pressure terms), so that $T_{00} -{1\over
  2} \,g_{00} T$ can be replaced by $\rho \,c^2/2$. Then only the $g_{00}$
part of the metric has material sources. Its weak field equation is 
\begin{equation}
\partial^2 g_{00} = -{8\pi \,G\over c^2}\,\Bigl(\,
\rho -\,{c^2\,\Lambda \,g_{00}\over 4\pi \,G}\,\Bigr)\,.
\label{eq:poissong00}\end{equation}

As $\partial^2 g_{00} =\partial^2 h_{00}$ because the derivatives
of $\eta_{00}=1$ vanish, the weak-field equation is usually written in
a form where $g_{00}$ on the left-hand side is replaced by $h_{00}$,
and $g_{00}$ on the right-hand side is replaced by unity, since both
$h_{00}$ and $\Lambda$ are small quantities. However, it is
essential to retain the $h_{00}$ contribution to the
$\Lambda$ term on the
right-hand side, because it is the source of 
vacuum polarization effects that will give us a wave equation for the
potential. Therefore we need to write the weak field equation in terms of
$g_{00}$, not in terms of $h_{00}$. 

Since the small $\Lambda h_{00}$ term has generally been neglected
in previous literature for the weak-field case, let us explain why it
is essential here. $\Lambda\sim 1/r^2_\Lambda$, where $r_\Lambda$ is a
characteristic length scale (cf. Eq.~(\ref{eq:krs})). As we will
see in Sect.~\ref{sec:scalehelm}, the small observationally
determined value of $\Lambda$ corresponds to an $r_\Lambda$ that is
approximately equal to the radius of the particle horizon of the
universe. On distance scales $r\ll r_\Lambda$ one can safely ignore
the $\Lambda h_{00}$ term, as has been correctly done for instance in
\citet{stenflo-jetzer_sereno2006} when considering the effect of
$\Lambda$ on the dynamics of stellar systems. It is however incorrect
to neglect it for cosmological distances, when the condition $r\ll
r_\Lambda$ is no longer satisfied, because it is the sole source of
either exponential, Yukawa-like cutoff (when the cosmological constant is negative)
or wave behavior of the potential (when the cosmological constant is
positive) at the characteristic $r_\Lambda$ distance scale. 

As usual the identification $h_{00}=2\Phi$ is made to
satisfy the Newtonian limit, where $\Phi$ is the gravitational
potential, here per unit energy (not per unit mass, which would differ by
the factor $1/c^2$), so that 
\begin{equation}
g_{00} =1+2\Phi\,.
\label{eq:g00}\end{equation}
The stationary version of Eq.~(\ref{eq:poissong00}) then gives us an
extended Poisson equation with $\Lambda$ term
for the potential $\Phi$: 
\begin{equation}
\nabla^2 \Phi +2\Lambda\,\Phi= {4\pi\, G\over c^2}\,(\,
\rho -\,2\rho_\Lambda\,)\,.
\label{eq:poissonlambda}\end{equation}
Apart from the definition of $\Phi$ with respect to unit energy instead of unit
mass, the equation would be identical to Eq.~(\ref{eq:poissuniv}), if it
were not for the profound $2\Lambda$ term on the left-hand side,
which represents a feedback of the medium (the vacuum) to the
potential $\Phi$. 

Equation (\ref{eq:poissonlambda}) explicitly brings out the two
physical roles played by the cosmological constant $\Lambda$: (1) The
$\rho_\Lambda$ term on the right-hand side is a source of repulsive
gravity, while (2) the $\Lambda$ term on the left-hand side provides
feedback to the potential, similar to the effect of vacuum
polarization. It is important to remember that these two roles reflect
two faces of the same coin, namely the two terms that make up $g_{00}$
in Eq.~(\ref{eq:g00}). Role (1) comes from the term 1, role (2)
from the term $2\Phi$. Both terms always contribute in concert. We do not
have the freedom to change their relative proportions. 

This unity of the two roles is implicitly contained in 
Eq.~(\ref{eq:poissong00}), which can be rewritten in a form that makes its
structural similarity to field theory formulations transparent: 
\begin{equation}
\partial^2 \varphi \,-\,m^2\,\varphi\,-J= 0\,,
\label{eq:kleingordon}\end{equation}
where
\begin{eqnarray}\label{eq:phijk2}
\varphi &\equiv& g_{00}\,,\nonumber\\ J &\equiv&\!\!\!-\,{8\pi\, G\over c^2}\,\rho
\,,\\m^2\!&\equiv&2\Lambda\,.\nonumber 
\end{eqnarray}

In general the gravitational field is a tensor field, while
Eq.~(\ref{eq:kleingordon}) represents it as a scalar field, because we
have disregarded the off-diagonal components of the stress-energy
tensor. This simplification is valid for our exploration of the
gravitational potential, and in particular when we later consider
global wave modes in a
cosmological medium that is isotropic and homogeneous. 

With this formulation the effect of $\Lambda$ appears
exclusively in the single $m^2$ term, there is no separate vacuum
energy density $\rho_\Lambda$ that combines with the mass density
$\rho$. This unification occurs because $\varphi$ is a physical field
that represents the metric (more precisely its $g_{00}$ component),
which implicitly contains both $\Lambda$ effects, in contrast to the
potential $\Phi$, which only represents a fractional aspect of the metric. 

In standard QFT (quantum field theory) $m$ represents a mass scale, while
here it is more convenient to let it represent a wave number $k_\Lambda$. 
This difference is however immaterial and only dependent on the 
choice of dimensions that we use for $\varphi$. Equation 
(\ref{eq:kleingordon}) then has formal similarity to the
Klein-Gordon equation, except for the sign of the $m^2$ term. The
Klein-Gordon sign is the origin of the Yukawa-type exponential cutoff
of the potential. The solution is the same if we formally replace the
Yukawa mass $m$ with the imaginary mass $i\,m$. This leads to a
potential with oscillatory behavior. 

If the $\Lambda$ term in the Poisson Eq.~(\ref{eq:poissonlambda})
were negative, the equation would be a so-called screened Poisson
equation and give rise to a Yukawa potential. In reality, however, the
term is positive, which gives us a Helmholtz equation with oscillatory
solutions \citep[cf.][]{stenflo-roza2017}. Such wave equations are
familiar in numerous areas of physics. The Schr\"odinger equation
belongs to this type. 

It is helpful to represent the wave number in terms of a
characteristic distance scale for the oscillations: 
\begin{equation}
m\equiv k_\Lambda \equiv 2\pi/r_\Lambda
\label{eq:krs}\end{equation}
It follows that $r_\Lambda = 2\pi/\sqrt{\,2\Lambda}\,$. It turns out
to be equal to the causal radius of the universe (distance to
the particle horizon), as we will see in Sect.~\ref{sec:origin}
(cf. Eq.~(\ref{eq:m2derived})). 

The scalar Eq.~(\ref{eq:kleingordon}) follows from the Lagrangian 
\begin{equation}
\mathcal{L} ={\textstyle{1\over
    2}}\,[\,(\partial^2\varphi)^2\,+\,m^2\,\varphi^2\,]\,+\,J\,\varphi\,. 
\label{eq:lagrange}\end{equation}
Its Green's function or propagator $D(x-y)$ is determined by 
\begin{equation}
-(\partial^2\,-\,m^2)\,D(x-y)= \delta^4(x-y)
\label{eq:propeq}\end{equation}
with the solution 
\begin{equation}
D(x-y)= \int\,{{\rm d}^4 k\over (2\pi)^4}\,{e^{ik(x-y)}\over k^2 + m^2}\,.
\label{eq:propsol}\end{equation}
We will refer to the expression for the propagator in Sect.~\ref{sec:indosc}.


\subsection{Feedback from the vacuum to the gravitational force
  field}\label{sec:vaceffects}  
The stationary version of Eq.~(\ref{eq:poissong00}) is the
Helmholtz equation 
\begin{equation}
\nabla^2 g_{00} +k_\Lambda^2 \,g_{00} ={8\pi\, G\over c^2}\,\rho\,,
\label{eq:helmstatg00}\end{equation}
where $k_\Lambda^2\equiv 2\Lambda\equiv (16\pi\, G/c^2)\,\rho_\Lambda$ with our
previous definitions. $\rho$ is the mass density, while all vacuum
effects come from the $k_\Lambda^2$ term. Below we will give the
solution of the equation for a point source. In this case we need to 
replace $\rho$ in the source term on the right-hand
side with $M\,\delta^3(r)$, where $M$ is the mass of the point source,
and $\delta^3(r)$ is the 3D Dirac delta function. 

The left-hand side of Eq.~(\ref{eq:helmstatg00}) has the structure
of a negative Hamiltonian in the time-independent Schr\"odinger
equation, when the force field is attractive so that the potential
energy is negative. The $k_\Lambda^2$ term corresponds to minus the
potential energy in the Schr\"odinger equation if 
$g_{00}$ plays the role of the wave function. This analogy illustrates why
we get a wave behavior for $g_{00}$ that is similar to waves in quantum physics. 

The general point source solution of Eq.~(\ref{eq:helmstatg00}) in
spherical coordinates is 
\begin{equation}
g_{00}=\sum_{\ell =0}^\infty\,\sum_{m=-\ell}^\ell\,[\,a_{\ell
  m}\,j_\ell(k_\Lambda r)
+b_{\ell m}\,y_\ell(k_\Lambda r)\,]\,Y_\ell^m(\theta,\varphi)\,,
\label{eq:generalsol}\end{equation}
where $j_\ell$ and $y_\ell$ are the two orthogonal
spherical Bessel functions of 
order $\ell$ (the radial wave functions in quantum mechanics), and
$Y_\ell^m$ are the spherical harmonics. The coefficients are
determined by boundary conditions. 

While the non-radial solutions lead to the rich structuring that we
encounter in atomic
physics, we assume that the gravitational potential exhibits exact
spherical symmetry. This implies that 
$\ell,\,m=0$, so that only the spherical Bessel functions $j_0$ and
$y_0$ need to be considered. 
\begin{eqnarray}
j_0&=&\phantom{-\,}{\sin k_\Lambda r\over r}\,,\nonumber\\ y_0&=&
-\,{\cos k_\Lambda r\over r}\,.
\label{eq:j0y0}\end{eqnarray}
 The solution then reduces to 
\begin{eqnarray}
g_{00}=a_{00}\,j_0 +b_{00}\,y_0\,.
\label{eq:reducedsol}\end{eqnarray}

Let us here denote the gravitational potential from a point source by
$\phi$ to distinguish it from $\Phi$, which referred to the total potential that included extended
sources. According to Eq.~(\ref{eq:g00})
\begin{eqnarray}
\phi={\textstyle{1\over 2}}\,(g_{00}-1)\,.
\label{eq:phig00}\end{eqnarray}
Let $\phi_N$ refer to the
corresponding Newtonian potential, which represents the case when
$\Lambda =0$. The Newtonian limit is expressed through  
\begin{eqnarray}
\phi =\phi_N \,\,\,{\rm when}\,\,r\ll r_\Lambda\,,
\label{eq:newtlim}\end{eqnarray}
which represents the inner boundary condition for 
Eq.~(\ref{eq:reducedsol}). 
\begin{eqnarray}
\phi_N =-\,{\textstyle{1\over 2}}\,{r_b\over r}
\label{eq:phin}\end{eqnarray}
with the definition 
\begin{eqnarray}
r_b \equiv {2GM\over c^2}\,.
\label{eq:rb}\end{eqnarray}
We recognize this as the radius of the event horizon for a black hole
of mass $M$. Here we use it to define the convenient parameter $r_b$, 
using the symbol $\equiv$ instead of $=$ to make it clear that we are
not suggesting that we are dealing with solutions for actual black
holes. In gravitational physics the same expression appears in
different contexts. For instance, for an
infinite, homogeneous universe with flat metric, 
Eq.~(\ref{eq:rb}) is valid when we let $r_b$ represent the Hubble radius
and $M$ the total mass inside the Hubble radius. 

The inner boundary condition Eq.~(\ref{eq:newtlim}) then gives us 
\begin{eqnarray}
b_{00} =r_b \,.
\label{eq:innerbc}\end{eqnarray}
The remaining parameters ($k_\Lambda$ and $a_{00}$) need to be
determined by outer boundary conditions, which we will identify later
(Eqs.~(\ref{eq:m2derived}) and (\ref{eq:betaatbound}) in 
Sect.~\ref{sec:origin}, with the use of the definition in 
Eq.~(\ref{eq:betadef})). 

Empty space or the vacuum state refers to the case when
$\rho =0$. Let us denote the gravitational potential of the vacuum
state by $\Phi_\Lambda$. It then follows from 
Eq.~(\ref{eq:phig00}) that 
\begin{eqnarray}
\Phi_\Lambda =-\,{\textstyle{1\over 2}}\,,
\label{eq:phivacuum}\end{eqnarray}
because $g_{00}\to 0$ as $1/r$ according to 
Eq.~(\ref{eq:reducedsol}), when the distance $r$ from 
the physical sources goes to infinity. This is an intriguing result,
since the number ${\textstyle{1\over 2}}$ is familiar from quantum
physics as the energy of the vacuum state. There it arises from the
quantization of the harmonic oscillator. In contrast, no quantization
condition has been applied here, although the Helmholtz 
Eq.~(\ref{eq:helmstatg00}) has structural similarity to the equation for a
harmonic oscillator. In our context the non-zero, flat $\Phi_\Lambda$ arises
because the gravitational field, which is represented by the metric
and not by the potential alone, is a superposition of the potential (the
fluctuating part) and the flat Minkowski background. The gravitational
field that is represented by $g_{00}$ does 
not have a non-zero vacuum state. The minus sign of $\Phi_\Lambda$ is
related to the circumstance that gravitation is attractive for
``charges'' (masses) of the same sign. The distinction between $g_{00}$ and
$2\phi$ vanishes when we consider the gravitational forces or
accelerations, because the derivative of the flat part of the
potential is zero. 

The gravitational acceleration is obtained from the potential through $g_{\rm acc}=
-\,\partial\phi/\partial r =-\,{\textstyle{1\over 2}}\,\partial
  g_{00}/\partial r$, which for the Newtonian case is $g_N 
=-r_b/(2r^2)$ according to Eq.~(\ref{eq:phin}). Here we add index
``acc'' to $g$ to make clear that we 
are talking about the acceleration and not the metric, which is also
referred to by $g$. From Eqs.~(\ref{eq:reducedsol}),
(\ref{eq:phig00}), and (\ref{eq:innerbc}) we find the ratio between
the Helmholtz and Newtonian accelerations to be  
\begin{equation}
g_{\rm acc}/g_N =(1+\beta\,x)\cos x\, +\,(x-\beta\,)\sin x\,,
\label{eq:gratio}\end{equation}
where for convenience we have introduced the dimensionless distance
scale 
\begin{equation}
x\equiv k_\Lambda r
\label{eq:xdefklr}\end{equation}
and the simplified notation 
\begin{equation}
\beta \equiv a_{00}/b_{00}\,.
\label{eq:betadef}\end{equation}
The value of $\beta$ needs to be fixed by an outer boundary condition,
which we will address in Sect.~\ref{sec:bccausal}.


\subsection{Scale of the Helmholtz oscillations}\label{sec:scalehelm} 
As we saw in Eq.~(\ref{eq:krs}), the scale $r_\Lambda$ of the
Helmholtz oscillations, defined by $k_\Lambda r_\Lambda \equiv 2\pi$, is
$r_\Lambda =2\pi/\sqrt{\,2\Lambda}\,$. We now want to express it in units of
the Hubble
radius of the universe, $r_H =c/H_0$, where $H_0$ is the current value
of the Hubble constant. 
For a Friedmann universe with zero curvature
$H_0$ is related to the critical density $\rho_c$ through 
\begin{equation}
\rho_c={3H_0^2\over 8\pi\, G}\,.
\label{eq:rhoc}\end{equation}
In the standard cosmological models the cosmological constant
represents a vacuum energy density $\rho_\Lambda =\Omega_\Lambda \,\rho_c$,
where $\Omega_\Lambda=0.69$ according to 
\citet{stenflo-planck2016}. With Eq.~(\ref{eq:rholam}) we then
find 
\begin{equation}
{r_\Lambda\over r_H}={2\pi\over \sqrt{\,6\,\Omega_\Lambda}}\,,
\label{eq:rlamh}\end{equation}
which equals 3.1 if we insert the observationally determined value
for $\Omega_\Lambda$. As we will see in Sect.~\ref{sec:horizons} and
in Eq.~(\ref{eq:omlfinal}), $r_\Lambda$ is 
equal to the causal radius (distance to the particle horizon) of the
universe. It is the scale that is relevant 
to dark energy while being vastly larger than the scales that are
relevant to dark matter. Still the resolution of the dark energy
enigma leads to non-standard effects that have direct implications for
the nature of dark matter, namely that all of dark matter appears to
be baryonic, without the need for yet to be 
discovered exotic forms of matter. We will return to this issue in
Sect.~\ref{sec:darkmatter}.


\section{Origin and nature of the dark energy}\label{sec:origin}  
Throughout the previous section we have tried to clarify some of the
aspects in which gravity is affected by the $\Lambda$ term, when it is
inserted ad hoc in the Einstein equation in the form of a cosmological
constant. We have highlighted, in particular with the help of
Eqs.~(\ref{eq:kleingordon}),  (\ref{eq:phijk2}), 
and (\ref{eq:helmstatg00}), the fundamentally different roles played
by the $\Lambda$ term and the real physical sources (which are
represented by the density $\rho$). The role 
of the $\Lambda$ term is exclusively to generate a feedback of the vacuum
from the metric (or the gravitational field) to itself, and not as a
direct source of gravity, a role that is instead played by real 
matter-energy  (e.g. as demonstrated by 
Eq.~(\ref{eq:helmstatg00})). 

If the $\Lambda$ term would represent a physical field as a constant
to satisfy the energy-momentum conservation equation and the Bianchi
identities, then the near coincidence
of the $\Lambda$-induced wave scale with the 
size of the current cosmic horizon would be extraordinarily unnatural
as it would violate the Copernican principle (which asserts that we are not
privileged observers). The horizon scale has increased by
about 60 orders of magnitude since the Planck era, while $\Lambda$
would not change if it were a true constant. This cosmic coincidence
would not disappear unless we abandon the view that $\Lambda$ is 
part of the underlying equations. The alternative is that the 
$\Lambda$ effects instead emerge from boundary conditions
that constrain the solutions of the equations, as we will show
below. If we would try to account for such boundary conditions by
inserting a fitting parameter $\Lambda$ into the original
equations, this parameter would masquerade as if it were a
new physical field, which it is not. 

In the following we will show that the $\Lambda$ effects are induced
as a consequence of the finite age of the universe. This implies that
the wave scale that is represented by $\Lambda$ and the scale of the
causal horizon must remain linked 
throughout all epochs of cosmic history. The cosmic coincidence problem
then goes away.


\subsection{Vacuum energy induced by the finite age of the
  universe}\label{sec:vacenergy}   
Usually time is viewed as a coordinate along an infinite axis, which
extends backwards before the Big Bang and forwards into the (as yet
non-existing) future beyond the
present moment. If however the age of the universe is finite, time is bounded in a
physical sense, because it does not exist beyond the two temporal
boundaries (Big Bang and the present moment), as we will further
clarify below. Let $\Delta t$ denote
the length of time between these two boundaries. While a Fourier
decomposition of a constant field 
along an infinite time line would give a delta function $\delta
(\omega)$, an infinitely sharp peak at zero frequency, the
corresponding decomposition for a truncated time line would give a
frequency spread $\Delta\omega$ that is approximately given by 
\begin{equation}
\Delta\omega\,\Delta t\approx {\textstyle{1\over 2}}\,.
\label{eq:domegdt}\end{equation}
If we multiply both sides by $\hbar$ we recognize Heisenberg's
uncertainty principle, with $\Delta E=\hbar\,\Delta\omega$ interpreted as the
vacuum energy that is induced by constraining the temporal
interval. The smaller this interval is, the more violent are the
fluctuations in energy. These general arguments will be made more precise in
Sect.~\ref{sec:indosc}, where we develop a concrete, quantitative theory for the
emergence of a vacuum energy $\Lambda$ of the observed magnitude, as a
consequence of the finite age of the universe. 

If a finite-age universe is to be compatible with quantum mechanics
(and obey the Heisenberg uncertainty principle), then it is inevitable
that the universe must have started with a hot Big Bang, because the
energy variance $\Delta E$ that represents the vacuum fluctuations
goes to infinity when $\Delta t$ goes to
zero. Likewise $\Delta E$ goes to zero in the distant future because it is
inversely proportional to the age of the universe. This vacuum energy
density, induced by the existence of temporal boundaries or horizons,
represents a scale that is always linked to the horizon scale for any
magnitude of the horizon radius. 

While it is generally accepted that time may have a beginning 
  or edge at the Big Bang, time is usually considered to be 
  unbounded in the forward, future direction. This seems to contradict
  our notion of a finite time line that ends at the present moment and
  does not extend into the future, and which therefore represents a 1D
  temporal cavity. The reason why time is finite and not semi-infinite
  is that the future is unobservable. 

When we look out into the universe, we look back in time,
until we reach the edge presented by the Big Bang. One speaks of
look-back time, but there is no such thing as look-forward time. There
can be no causal influences reaching us 
from the future. Therefore the present represents a causal boundary
that cannot be crossed from the ``outside'' (the future). Our temporal 1D cavity is
constrained between two causal boundaries, at the Big Bang and at the
present time. Similarly, space is bounded, between ``here'' and the causal (particle)
horizon. The bounded, observable 4D volume increases continually as
the horizon advances both spatially and temporally (into the future) with the
age of the universe.


\subsection{Flatness of space}\label{sec:flatness}  
Equation (\ref{eq:helmstatg00}) explicitly showed us that only
real mass-energy (represented by $\rho$) but not $\Lambda$ is a source of
gravity. To understand this statement better, a
comparison with Debye screening may be helpful. The source of the
electric field is the point charge that is embedded in a plasma environment that is
filled with a continuum of charges. These environmental charges are
however not a source of the field, their role is to provide the screening of the
potential. Similarly, $\rho$ is the source, while $\Lambda$ provides
the feedback, which because of its sign does not result in screening, but 
instead endows the (Helmholtz) potential with wave properties. 

Gravity manifests itself by changing the metric. In the absence of
sources for gravity spacetime is flat. An empty universe 
($\rho =0$) has no gravity, no curvature. In a non-empty universe that expands forever
$\rho\to 0$ as time $t\to\infty$. The obvious boundary condition at infinity
for such a universe must therefore be that it has Minkowski metric and
not something else (like de Sitter metric). 

If the curvature of space is zero at a given epoch, then it follows
from the standard cosmological equations that it is zero at all
epochs. There is no need to invent a hypothetical inflationary phase
for the purpose of explaining why the metric of the universe is observed to be
flat. Since $\Lambda$ without $\rho$ cannot be a source of gravity, as
we have shown above, it follows that there cannot be an
exponentially expanding phase in the distant future, the end phase of
the universe is not a de Sitter phase. 

The starting point of the following treatment is therefore to consider
a homogeneous and isotropic universe with zero curvature and without
any cosmological constant present in the underlying equations. We will
then show how the finite age of the universe leads to effects, which are the same as
those of a cosmological constant $\Lambda$
with a value that matches the observed one. No free parameters
or model fitting are used to derive this. Nature did not have a choice
to generate $\Lambda$ effects of different magnitude.


\subsection{Cosmic horizons and conformal coordinates}\label{sec:horizons}  
In Sect.~\ref{sec:vacenergy} we gave qualitative arguments why
$\Lambda$-like 
effects should emerge when we constrain the time line, because
the set of Fourier modes that can fit within the finite interval gets
restricted. To express these ideas in a precise and quantitative way we first
need to clarify the meaning of causal boundaries or horizons in an
expanding universe. This also allows us to clarify what we mean by the
stationary solution for the Helmholtz potential in
Sect.~\ref{sec:vaceffects}, when we are dealing with an expanding
universe with a time-varying scale factor $a(t)$. 

Observations tell us that the current age of the universe is
13.8\,Gyr, while the Hubble time ($1/H_0$) is 14.4\,Gyr. Although the
Hubble radius at 14.4 billion light years (GLyr) is often referred to
as the ``cosmic horizon'', this terminology is a bit misleading,
because the Hubble radius is ``just'' a parameter, not a physical
horizon. The causal or particle horizon, from beyond which no forces or interactions
can reach us, is at 46.9\,GLyr. This distance equals the speed of
light times the {\it  conformal} age of the universe (which is
46.9\,Gyr). It is the time it would take for a photon to reach us from
the causal horizon if the
universe would stop expanding. The normal and the conformal age of the
universe are different because the points from which the propagating
light tries to reach us continually recede from us due to the cosmic
expansion. 

The radius of the horizon is differentially stretched by the
expansion, as expressed by the distance-redshift relation. The usual
way to describe this is in terms of the Robertson-Walker metric with a
scale factor $a(t)$ that multiplies the spatial differential distance. For
many purposes it is much more useful to instead use conformal
coordinates, with which the scale factor has been transformed away so
that the metric formally becomes Minkowski-like. 

While ordinary cosmic time $t=\int_0^t {\rm d}t$, conformal time
$\tau$ is defined by 
\begin{equation}
\tau =\int_0^t{{\rm d}t\over a}
\label{eq:conftime}\end{equation}
or, equivalently, ${\rm d}\tau/{\rm d}t=1/a$. In a spacetime diagram
based on conformal coordinates (conformal time vs. comoving radial
distance), light rays are straight lines with a slope given by the
speed of light, just like they are in inertial coordinate systems. 

The conformal coordinate background provides an arena in
which it makes sense to make Fourier decompositions, because conformal
invariance implies that geometric shapes and angles are preserved. The
sine and cosine components of a Fourier decomposition do not get
deformed (differentially redshifted) by the stretching of space when we use 
conformal coordinates, they retain their sine and cosine
shapes. It is against this coordinate background that
one should interpret the metric perturbations that one finds when
solving the weak-field Einstein equations for the stationary case, as
we did in Sect.~\ref{sec:standder}. In a normal,
expanding coordinate system it is not well defined what we mean by a
stationary case. The Helmholtz potential that we found in
Sect.~\ref{sec:standder} needs to be expressed in terms of conformal
coordinates.


\subsection{Euclidian spacetime}\label{sec:euclid}
The treatment of the $\Lambda$-like 
effects that emerge because the observable time line is constrained
can best be done in 4D Euclidian 
spacetime, because time can then be viewed as an angular
coordinate, which reveals the existence of a resonance condition that 
is the reason for the effects that we refer to as dark energy. The transformation
from Minkowski to Euclidian spacetime is done through 
Wick rotation in the complex plane, such that ordinary time $t$
becomes Euclidian time $t_E= i\,c\,t$, where we have inserted $c$ to
express $t_E$ in spatial units like the three other coordinates. Since
time now becomes imaginary, it may be interpreted as an angular
coordinate with period $2\pi$, which corresponds to the length $\ell$
of a finite time string that has periodic
boundary conditions. In our case $\ell$ is the conformal age of the universe 
expressed in Euclidian time.  The various wave modes of the Fourier
decomposition then have wave number $2\pi\,n/\ell$, where $n$ is an
integer. 

The transformation to Euclidian spacetime leads to remarkable, even
miraculuos advantages and insights. The Lagrangian, which is used for
the formulation of the Einstein equations (cf. 
Eq.~(\ref{eq:lagrange}) for the weak-field case), becomes the Hamiltonian,
which is the agent that drives the cosmic evolution. Quantum
field theory QFT transforms into the structure of classical 
statistical mechanics. The path integral in field theory then corresponds to
the partition function in statistical mechanics, with
the oscillating phase factors in QFT now appearing as the Boltzmann
factors, which allow the definition of a temperature. The
transformation thereby establishes a direct link between 
field theories like general relativity or QFT and thermodynamics. In
particular it provides
a direct route to the derivation of the Hawking temperature of black
holes. For a brief introduction to this topic, see for instance
\citet{stenflo-zee2010}.


\subsection{Wave modes induced by the finite age of the
  universe}\label{sec:indosc} 
The following wave mode discussion will relate to the treatment of the
weak-field approximation that we did in Sect.~\ref{sec:standder}. This
approximation is valid for all cosmological epochs except for the very
early universe, in particular when we approach the Planck era. However,
as this strong-field era is of miniscule temporal extent as compared
with the relevant cosmological time scales that we are dealing with
here, the resonant condition that we will identify as the origin of
the $\Lambda$ effects remains valid although it is based on the weak-field
treatment. 

In the Euclidian spacetime the 4D d'Alembertian operator becomes
$\partial^2=\partial^2/\partial t_E^2 +\nabla^2$, because this
spacetime has signature $(+++\,+\,)$. Its inverse, representing the field
propagator, is $\sim 1/k^2$, where the square of the 4D wave number
can be written as $k^2=k_4^2 +\vec{k}^2$. Here
$\vec{k}^2=k_1^2+k_2^2+k_3^2$, with $k_{1,2,3}$ representing the usual
spatial wave numbers, while $k_4$ is now the angular frequency of
Euclidian time and represents the temporal modes. 

As a consequence of the periodic boundary conditions (that result
because Euclidian time is cyclic), the temporal Fourier transform with
the factor $\exp(i\,k_4 t_E)$ gets restricted to values for which $k_4
=2n\,\pi/\ell$ for the $n$th harmonic, where the string length 
$\ell=i\,c\,t_c$, with $t_c$ being the conformal age of the
universe, 46.9\,Gyr according to standard cosmology. It follows that
the square of the 4D wave number  
is $k^2=\vec{k}^2+(2n\, \pi/\ell)^2$ for the $n$th harmonic mode,
implying that we have effectively lost one dimension. The expression
represents a discrete set of stationary modes for the spatial 3D wave
number.  Because it is 
the temporal dimension that has been lost, we have retrieved the
stationary case that we need to make direct comparison with the
corresponding stationary case of Eqs.~(\ref{eq:propsol}) and
(\ref{eq:helmstatg00}). This allows us to 
relate our expression for $k_4$ with the dark energy parameters 
$\Lambda$ or $k_\Lambda$. 

The partition function that governs the probability distribution
over the possible wave modes is the sum over the
respective Boltzmann factors that are generated by the Wick rotation:
$\sum_{n=1}^\infty e^{-2n\pi} =1/(e^{2\pi} -1)$, which shows how a Planck
distribution emerges. As the probability for excitation to the next higher
harmonic decreases by the Boltzmann factor $e^{-2\pi}\approx 1/535$, it
is a good approximation to only consider the fundamental mode (with
$n=1$) as relevant. We will do this here. 

With the help of the Boltzmann factor one may introduce a
temperature. Although we do not
need to make use of the temperature concept in order to derive
$\Lambda$, we mention it here because it may be of interest to 
indicate how it is related to the Hawking temperature and a
horizon. If we in the Boltzmann factors make the identification
$\hbar\,\omega/(k_B T)=\omega\, t_c$, we get 
$T=\hbar/(k_B t_c)$, which here has the stupendously small value of
about $10^{-29}$\,K due to the gigantic value of $t_c$. Using the
expression for the Schwarzschild radius of black holes, $r_{\rm
  BH}\equiv c\,t_{\rm BH}=2GM/c^2$, where we for convenience of comparison
have introduced a black hole time scale $t_{\rm BH}$, we can convert
the standard expression for the Hawking temperature $T_H$ to the form
$T_H=\hbar/(4\pi k_B t_{\rm BH})$. This expression is the same as that of our
simplistic derivation if we replace $t_{\rm BH}$ with $t_c$, with the
exception of the numerical factor 
$4\pi$, which may be due to the greatly different geometrical
situations in the two cases (as our universe does not have a
Schwarzschild metric). While not directly needed for our
derivation of $\Lambda$, it is worth paying attention to the potentially
profound implicit connections that causal horizons have with
thermodynamics and quantum physics. 

When converting back from Euclidian age $\ell$ to ordinary conformal
age $t_c$ while disregarding the higher harmonics, we get in conformal
coordinates $k^2=\vec{k}^2-m^2$, where 
\begin{equation}
m^2 \equiv k_\Lambda^2=[\,2\pi/(c\,t_c)\,]^2\,.
\label{eq:m2derived}\end{equation}
Here we have for later use (in Sect.~\ref{sec:bccausal}) reintroduced the wave number
$k_\Lambda$ that we first introduced in Eq.~(\ref{eq:krs}). 
The field propagator $1/k^2$ that we started off with has thus become
$1/(\vec{k}^2-m^2)$, exclusively as a result of the finite length of
the time line. As
this now represents a stationary spatial wave pattern, it needs to be compared
with the propagator of Eq.~(\ref{eq:propsol}) for the stationary
case, when $\partial/\partial t$ and the corresponding wave number
$k_0$ are zero. In this case the denominator in 
Eq.~(\ref{eq:propsol}) is $-\vec{k}^2+m^2$, which is identical to the
propagator that we derived via bounded time, except for the overall
sign. This global sign is however immaterial. It is a consequence of
using the signature ($+--\,-$) for the Minkowski metric rather than ($-++\,+$), and
because of the circumstance that when we transformed to Euclidian
coordinates, we switched the 
sign of the temporal but not the spatial part in the signature of the
metric. While the overall sign does not matter, the relative sign
between the $\vec{k}^2$ and $m^2$ terms is essential. It agrees with
Eq.~(\ref{eq:propsol}), which implies that the stationary
gravitational potential is of Helmholtz and not Yukawa type, and that
the cosmological $\Lambda$ parameter, which is obtained from the
identification $m^2 =2\Lambda$ of Eq.~(\ref{eq:phijk2}), is
positive. 

For readers who may be confused by this derivation of the sign for
$\Lambda$, because we have gone back and forth between Euclidian and
Minkowski coordinates, the following heuristic argument why the sign
of $\Lambda$ must be positive may be helpful. The finite time string
may be viewed as an infinite time line on which we have imposed a
rectangular window of width $t_c$, which cuts off everything outside
the window. This rectangular restriction is qualitatively similar to
the exponential Yukawa-type cutoff, when applied to the time line. In
the time domain this cutoff has the consequence that $k_0^2$ changes 
to $k_0^2 +m^2$ (with $m$ representing the inverse cutoff scale), where
we notice the same signs in front of $k_0^2$ and $m^2$. However, this
cutoff-induced $m^2$ term then has a sign that is opposite to that of the
spatial $\vec{k}^2$ term because of the signature of the Minkowski
metric. As we have shown before, this has the consequence that in the
spatial domain the gravitational potential is of Helmholtz and not
Yukawa type, and that $\Lambda$ is positive. 

Let us note that for a homogeneous universe without spatial gradients,
the circumstance that $k_0^2$  combines with $m^2$ with a + sign
implies a Yukawa-type exponential temporal behavior (induced by time
being finite). This exponential temporal behavior is
in cosmology usually expressed in terms of a de Sitter ($\Lambda$) term,
which, if being the sole source of evolution, would lead to an
exponential expansion, in contrast to exponential decay in the
standard Yukawa case. However, in both the de Sitter and Yukawa cases
the equations allow both exponentially 
growing and decaying solutions, the selection is made by the boundary
conditions that we impose. The de Sitter solutions are growing,
because we choose to start off with a compact Big Bang.


\subsection{Boundary condition at the causal
  horizon}\label{sec:bccausal}  
With the concepts and tools that we have developed in the preceding
sections, let us now come back to our exploration of the Helmholtz
potential that we did in Sect.~\ref{sec:gravpot}, since we are now in
a position to define the previously unspecified outer boundary
condition. This will allow us to illustrate (in
Fig.~\ref{fig:gacc} below) the behavior of Helmholtz gravity and to
compare it with Newtonian gravity. The treatment in the present section is
not needed for our derivation of the dark energy parameter
$\Omega_\Lambda$, which is the subject of the following 
section, it is only needed for the completion of 
our presentation of the behavior of Helmholtz gravity. 

The particle horizon constitutes the natural outer boundary, since no
causal effects or interactions emanating from objects beyond this
distance can reach us, including all gravitational effects. Then continuity demands
that all interactions, including all accelerations, must vanish at this distance. 

A treatment of the inner boundary condition, in the Big Bang at the beginning
of time, is beyond the scope of the present paper, because it is in
the realm of quantum gravity. Let us however briefly
reflect on the implications of the requirement that all interactions
should vanish at both causal boundaries, also the inner one, to
satisfy continuity. Similar to the 
outer boundary case, there should be no interactions from anything before the
beginning of time, because this part of the universe does not exist
(if physical time is truly finite). 

The vanishing of all interactions as time $t\to 0$ implies that the
beginning is a state of asymptotic freedom. At 
sufficiently small temporal and spatial scales the gravitational interactions should go to
zero if this natural boundary condition is to be satisfied. While we are not
yet in a position to specify the scales at which such asymptotic freedom
would be reached, it is reasonable to expect the transition
to be related to the scales of the Planck era, but this is a topic that
we will not pursue more here. 

According to Eq.~(\ref{eq:gratio}) the Helmholtz gravitational
acceleration $g_{\rm acc}$ can be written (ignoring the constant of
proportionality) as  
\begin{equation}
g_{\rm acc}\sim -[\,(1+\beta\,x)\cos x\, +\,(x-\beta\,)\sin x\,]\,/\,x^2\,.
\label{eq:gacc}\end{equation}
With the same constant of proportionality the corresponding Newtonian
acceleration is $\sim -1/x^2$. Here $x$ is the dimensionless distance
parameter defined by $x\equiv k_\Lambda r$ as in 
Eq.~(\ref{eq:xdefklr}), with the wave number given by
$k_\Lambda=2\pi/(c\,t_c)$ according to 
Eq.~(\ref{eq:m2derived}). As before $t_c$ is the conformal age of the
universe, and $r$ is the comoving distance coordinate (speed of light
times conformal time). 

$\beta$ is a parameter of the Helmholtz solution that has to be
determined by an outer boundary condition. This condition is that
$g_{\rm acc}$ must vanish at the causal boundary,
where $r=c\,t_c$ and therefore $x=2\pi$. While $\sin x$ then vanishes at the
boundary, $\cos x$ does not. Therefore $g_{\rm acc}$ can
only vanish there if $1+2\pi\,\beta=0$, which
unambiguously fixes the value of $\beta$:
\begin{equation}
\beta =-{1\over 2\pi} \approx -0.159\,.
\label{eq:betaatbound}\end{equation}

\begin{figure}
\resizebox{\hsize}{!}{\includegraphics{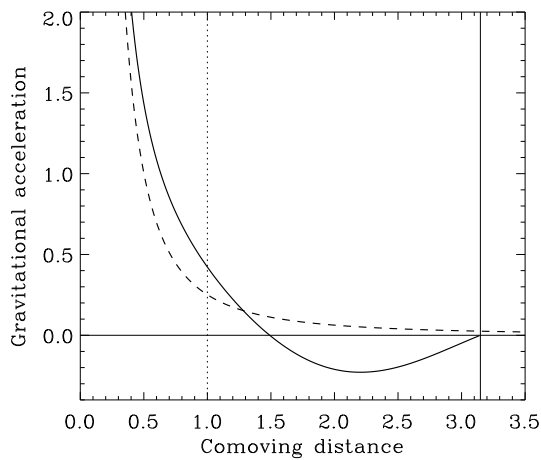}}
\caption{Gravitational accelerations for the Helmholtz (solid)
  and the Newtonian (dashed) cases as functions of the comoving radial
  distance in units of the Hubble radius (marked by the vertical
  dotted line). While the Newtonian curve ignores the boundary and
  only reaches zero at infinity, the Helmholtz curve ends at the
  causal boundary with zero acceleration. The location of the causal
  boundary (marked by the solid vertical line) is determined
  by Eq.~(\ref{eq:omlfinal}), which also uniquely determines the
  value of the dimensionless vacuum energy parameter
  $\Omega_\Lambda$. Note that the Helmholtz acceleration changes sign
  and becomes repulsive beyond the halfway point to the causal 
  horizon. 
}\label{fig:gacc}
\end{figure}

In Fig.~\ref{fig:gacc} we have used this value for $\beta$ to plot the
expression on the right-hand side of Eq.~(\ref{eq:gacc}) (after
reversing its sign) 
as the solid curve, with the corresponding Newtonian $1/x^2$ as the
dashed curve. The horizontal axis represents the comoving radial distance
in units of the Hubble radius, which is marked by the vertical dotted
line. The curve for the Helmholtz acceleration ends where it meets the
vertical solid line that marks the 
position of the causal horizon. The exact position of the causal
horizon in these distance units is uniquely determined from the
solution of Eq.~(\ref{eq:omlfinal}), as will be explained in the
next section. 

We notice that the attractive Helmholtz force is substantially
stronger than the Newtonian force at smaller and intermediate
distances. About halfways to the outer boundary the Helmholtz force changes
sign and becomes repulsive, before it has to vanish at the boundary. 

The enhancement of the force at smaller distances may raise
the question whether such a deviation from Newtonian gravity could
potentially be an explanation for the phenomenon behind what we refer
to as ``dark matter''. The answer to this question is ``no''. The
enhancement that we get with our Helmholtz force does not exceed a
factor of two, which is far too little to account for dark matter. To
be a viable explanation for dark matter the enhancement would need to
be larger by more than an order of magnitude and occur at scales about
$10^4$ times smaller than the horizon scale. We are not aware of any physical
effects that would lead to modified gravity with such properties.


\subsection{Link to the Planck era and the uniqueness of
  $\Omega_\Lambda$}\label{sec:linkplanck}  
The critical value $\rho_c$ of the mean mass density in a Friedmann
universe with zero spatial curvature is 
\begin{equation}
\rho_c ={3\over 8\pi \,G \,t_H^2}\,,
\label{eq:rhocrit}\end{equation}
where $t_H =1/H$ is the Hubble time. This scaling between $\rho$ and
time $t$ is identical to that for a black hole of radius $r_{\rm
  BH}\equiv c\,t_{\rm BH}$ (used to define the time scale $t_{\rm
  BH}$) and mass $M_{\rm BH}$. From the expression 
\begin{equation}
t_{\rm BH} ={2\,GM_{\rm BH}\over c^3}
\label{eq:tbh}\end{equation}
for the Schwarzschild radius and assuming a homogeneous density
distribution $\rho$ so that $M_{\rm BH}=4\pi\rho \,c^3\,t_{\rm BH}^3/3$,
we recover Eq.~(\ref{eq:rhocrit}) if we replace $\rho$
with $\rho_c$ and $t_{\rm BH}$ with $t_H$. 

In Friedmann cosmology $\rho_c$ marks the boundary case between open
and closed model universes. For black holes the corresponding density
marks the boundary between stability and instability with respect to black-hole
formation. To see how this relates to the Planck era, and subsequently
how the present mean density of the universe relates to the Planck
density, let us first define the Planck era as the epoch when mini
black holes get spontaneously formed by vacuum fluctuations. Let us
for convenience use the notation $\Delta E \equiv M_{\rm BH}\,c^2$ to
define the energy content of such a mini black hole, and denote
$\Delta t\equiv t_{\rm BH}$. Then, with the use of 
Eq.~(\ref{eq:tbh}), the Heisenberg criterion for 
spontaneous formation of such mini black holes is 
\begin{equation}
\Delta E \,\Delta t ={c^5\over 2\,G}\,\,t_{\rm BH}^2\,=\textstyle{1\over 2}\,\hbar\,,
\label{eq:spontform}\end{equation}
which defines the Planck time $t_P$ as the $t_{\rm BH}$ that satisfies
Eq.~(\ref{eq:spontform}): 
\begin{equation}
t_P=\left({\hbar \,G\over c^5}\right)^{\!1/2}\!.
\label{eq:tplanck}\end{equation}

Because all black holes and flat Friedmann universes, regardless of
their size, scale according to Eq.~(\ref{eq:rhocrit}), we can
relate the present critical mean density $\rho_c$ of the universe to
the density $\rho_P$ in the Planck era (at time $t=t_P$) through 
\begin{equation}
{\rho_c\over \rho_P}=\left({t_H\over
    t_P}\right)^{\!-2}=H_0^2\,t_P^2\,\approx 10^{-122}\,,
\label{eq:rhoscaling}\end{equation}
since $t_P=5.38\times 10^{-44}$\,s and $t_H=14.4$\,Gyr $=4.54\times
10^{17}$\,s. This beautiful scaling relation over 122 orders of
magnitude would be wrecked if a hypothetical inflationary phase would
be introduced. 

Let us now see how the emergent vacuum energy that is represented by
the $\Lambda$ parameter relates to all this. The most convenient
representation of $\Lambda$ is in the form of the dimensionless
parameter $\Omega_\Lambda$, which we introduced in
Sect.~\ref{sec:scalehelm}. It is defined via the vacuum energy density
$\rho_\Lambda$ that was introduced in Eq.~(\ref{eq:rholam})
through 
\begin{equation}
\Omega_\Lambda\,\rho_c\equiv \rho_\Lambda\equiv {c^2\Lambda\over 8\pi \,G}\,.
\label{eq:omldef}\end{equation}
With Eq.~(\ref{eq:rhocrit}) we then obtain 
\begin{equation}
\Lambda ={3\,\Omega_\Lambda\over c^2\,t_H^2}\,,
\label{eq:lamrhoc}\end{equation}
while from Eqs.~(\ref{eq:m2derived}) and (\ref{eq:phijk2}) we get
the same $\Lambda$ when derived as an emergent quantity that is a
consequence of the finite age of the universe: 
\begin{equation}
\Lambda ={2\,\pi^{\,2}\over c^2\,t_c^{\,2}}\,,
\label{eq:lamemergent}\end{equation}
where $t_c$ is the conformal age of the universe. Combining 
Eqs.~(\ref{eq:lamrhoc}) and (\ref{eq:lamemergent}) we find the expression
for the dimensionless $\Omega_\Lambda$: 
\begin{equation}
\Omega_\Lambda\,={2\over 3}\left({\pi\,t_H\over t_c}\right)^2.
\label{eq:omlfinal}\end{equation}

Note that $\Lambda$ in Eq.~(\ref{eq:lamemergent}) depends
  exclusively on the conformal age $t_c$, while $\Omega_\Lambda$ in
  Eq.~(\ref{eq:omlfinal}) depends on the ratio $t_c/t_H$. The
  reason why the Hubble time appears in Eq.~(\ref{eq:omlfinal})
  is that $\Omega_\Lambda$ represents the fraction of the critical
  density $\rho_c$ that is in the form of dark energy, and $\rho_c\sim
  1/t_H^2$ according to Eq.~(\ref{eq:rhocrit}). 

In the models of standard cosmology the ratio $t_c/t_H$ between the
conformal and Hubble times depends on the cosmological parameters,
including $\Omega_\Lambda$ (because it depends on the shape of the
$a(t)$ function), but it is close to $\pi$ for the parameters used in
standard cosmology. Equation (\ref{eq:omlfinal}) then gives
$\Omega_\Lambda\approx 
2/3$, which is consistent with the value adopted from observations. 

Due to the dependence of $t_c/t_H$ on $\Omega_\Lambda$, there is in
fact a unique solution for both $\Omega_\Lambda$ and $t_c/t_H$ from 
Eq.~(\ref{eq:omlfinal}), namely $\Omega_\Lambda=0.664$ and $t_c/t_H=3.15$ (in the case that
$\Omega_M$ is due to matter rather than radiation, see below). Any other value is
prohibited, since it would not be consistent with this equation. These 
values should be compared with the corresponding values derived by
\citet{stenflo-planck2016} from 
the CMB observations when applying the interpretational framework of
standard cosmology: 0.69 and 3.26 ($=46.9$\,GLyr / $14.4$\,GLyr),
respectively. The 
agreement between observations and theory can be considered good, in
particular since we do not use any free fitting parameters in our
theory, and the CMB observations have been interpreted with a theoretical
framework that is not identical to ours. 

From the definition of conformal time in Eq.~(\ref{eq:conftime}),
the solution for $t_c/t_H$ can be written as 
\begin{equation}
{t_c\over t_H} ={1\over t_H}\int_0^{t_U}{{\rm d}t\over a}=\int_0^\infty {{\rm d}z\over E(z)}\,,
\label{eq:tcthinteg}\end{equation}
where $z$ is the redshift, $t_U$ is the age of the universe in normal
cosmic time units, and 
\begin{equation}
E(z)=[\,\Omega_E\,(1+z)^n \,+\,\Omega_\Lambda\,]^{1/2}\,.
\label{eq:ez}\end{equation}
Since this expression represents the case of zero curvature, $\Omega_E=1-\Omega_\Lambda$
can be transformed away \citep[cf.][]{stenflo-longair2012}. Here we
have introduced parameters $\Omega_E$ and $n=3(1+w)$, where $w$ is the
equation of state parameter, to allow for the two
main cases when the universe is matter-dominated ($w=0$) as it is at present,
in which case $\Omega_E =\Omega_M$ and $n=3$, and when it is radiation
dominated ($w=1/3$ as in the early universe), in which case $\Omega_E
=\Omega_R$ (dimensionless density parameter for radiation) and
$n=4$. The expression for $t_c/t_H$ in Eq.~(\ref{eq:tcthinteg})
can be inserted in Eq.~(\ref{eq:omlfinal}), which can 
then be solved numerically. For the matter-dominated universe in which
we now live, $\Omega_\Lambda=0.66$, while in the radiation-dominated
era $\Omega_\Lambda=0.93$. Thus almost all of the energy
density in the early universe was in the form of the 
vacuum energy that is a consequence of the restricted extent of the physical time line.  

$\Omega_\Lambda$ remains constant for all times as long as there is no
change in the equation of state for $\Omega_E$. In the
radiation-dominated era  $\Omega_\Lambda$ stays at 93\,\%\ all the way
back to the Planck era, in the matter-dominated era it stays at
66\,\%\ for all future times. The present balance between $\Omega_M$ and
$\Omega_\Lambda$ leads to the value 1.07 for the ratio between the Hubble time and the age
of the universe (while it is $14.4/13.8=1.04$ according to
\citet{stenflo-planck2016}), which is close to unity. This 
implies that the expansion parameter $a(t)$ is nearly linear with
respect to time except for a small deceleration, and it will remain
so forever. There will be no 
transition to an exponentially expanding phase as is often believed
and which would result in an utterly empty universe. In the
radiation-dominated early universe the ratio Hubble time to age of the universe is slightly
smaller, about 96\,\%, which also implies
an almost linear expansion except for a small acceleration as a result of the larger
$\Omega_\Lambda$.

\section{Implications for dark matter}\label{sec:darkmatter}
The results of the previous section demonstrate how our explanation of dark energy as an
emergent phenomenon leads to non-standard features outside the
current framework, although it accounts for the observed $\Lambda$
effects and is largely consistent with overall aspects of standard
cosmology. The non-standard aspects represent an advantage, because
the theory becomes amenable 
to future observational tests that may bring in different
perspectives on unsolved issues in cosmology. 

Here we will focus on one such unsolved, long-standing enigma in
astrophysics: the nature of dark matter. There have been numerous
attempts to account for dark matter through 
a parametrized modification of Newtonian gravity to fit the observed
rotation curves of galaxies, starting with the MOND (modified
Newtonian dynamics) approach introduced by
\citet{stenflo-milgrom1983}. 

The term $a_{00}\,j_0$ in Eq.~(\ref{eq:reducedsol}) for the
Helmholtz potential actually has
the right functional form to give 
an excellent fit to the observed galaxy rotation curves if we were to
use $a_{00}$ and wave number $k_\Lambda \equiv
2\pi/r_\Lambda\equiv\sqrt{\,2\Lambda}$ 
(cf. Eqs.~(\ref{eq:krs}) and (\ref{eq:helmstatg00})) as free
parameters. However, in our 
case the wave scale implied by $k_\Lambda$ is not a free parameter but is
fixed by the value of $\Lambda$, which is observationally tied to the
accelerated expansion of the universe, and which is also theoretically
fixed via Eq.~(\ref{eq:omlfinal}). 

While our $k_\Lambda$ implies a
characteristic distance scale that is given by the radius of the
causal horizon, the
galaxy distance scale over which we need significant 
modifications of the Newtonian potential to account for the dark
matter signatures is smaller by at least about four orders of magnitude. Since
$\Lambda\sim 1/r_\Lambda^2$, we would need a value of $\Lambda$ that is
larger by a factor of about $10^8$ to induce significant Helmholtz
effects on the Newtonian potential at galaxy scales, but such a 
possibility is prohibited by cosmological constraints. 

There now exists rather convincing observational evidence that disfavor
an explanation of dark matter in terms of a modification of gravity,
in particular from the observations of a pair of colliding galaxy
clusters, the so-called Bullet Cluster 
\citep[cf.][]{stenflo-bullet2004,stenflo-bullet2006}, which reveal a
significant offset between the visible matter distribution and the
dark matter inferred from gravitational lensing. While this offset
implies that the invisible component must consist of collisionless
matter, it does not imply that this matter should be non-baryonic and
consist of some yet to be discovered weakly interacting massive
particles (WIMPs). 

The main reason why most of dark matter is believed to be non-baryonic
is the required compatibility with predictions from BBN (Big Bang
nucleosynthesis) calculations. A small baryonic mean density
($\Omega_B\approx 5$\,\%\ of the critical density $\rho_c$) is needed within the
standard cosmological framework for agreement of BBN with the
observed abundances of the light chemical elements. The situation however 
changes with our new theoretical framework, because the
expansion rate at the epoch of nucleosynthesis is different. Here we
will show that when this non-standard aspect is accounted for,
agreement between observed abundances and BBN calculations requires
that $\Omega_B$ is instead of the same order as the present $\Omega_M
=1-\Omega_\Lambda$, which implies that (within the 
uncertainties of the derivations) all of the dark matter is baryonic,
there may no longer be any justification for introducing anything exotic of not yet
discovered nature. 

For an overview of BBN physics we refer to
\citet{stenflo-peebles1993}. It is
outside the scope of the present paper to carry out 
comprehensive BBN calculations that fully account for the
non-standard effects of our new theoretical framework in quantitative
detail. We will limit our focus to the deuterium abundance, which is
the ``baryometer of choice'' \citep[cf.][]{stenflo-steigman2007}
because of its particularly high sensitivity to the adopted value of
the relative baryon abundance $\Omega_B$. 

The bottleneck for all subsequent BBN processes is deuterium
formation. There is competition between deuterium creation through
proton-neutron collisions and deuterium photodissociation by the
ambient radiation field. BBN could only start when the general photon energy 
fell below the binding energy of deuterium, which occurred when the
temperature in the Big Bang dropped below $10^9$\,K. Then all the free
neutrons got quickly captured to form deuterium. Once a significant
fraction of deuterium nuclei $d$ were available, they started to be used
up for the production of $^4$He, with $d+d$ collisions as the first
step, leading to tritium ($t$) or $^3$He formation, followed by $t+d$
and $^3$He$\,+\,d$ reactions. 

This had the result that all free neutrons ended up in $^4$He while
deuterium was destroyed in the process. The deuterium destruction was
however incomplete because of the density drop due to the
expansion of the universe, which brought the deuterium
destruction to a halt. The remaining, undestroyed abundance of
deuterium is therefore a very sensitive function of the initial baryon
density, parametrized by the dimensionless $\Omega_B$, and the
expansion rate that is responsible for cutting off the 
destruction process. This is the reason why deuterium is the
baryometer of choice, our measure of the $\Omega_B$ parameter. 

In standard cosmology the effects of spacetime curvature and a
hypothetical cosmological constant become negligible in the early
universe, because matter and radiation scale with the scale factor
$a(t)$ like $a^{-3}$ and $a^{-4}$ to become the dominant drivers of
the expansion in the early universe. The expansion rate gets
uniquely determined by the Friedmann solution that describes the
radiation-dominated era, for which the Hubble
time is 2.0 times the age of the universe. The only remaining free
parameter for the BBN calculations is $\Omega_B$, which is then
constrained to be about 5\,\%\ to agree with the observed deuterium
abundance. 

A non-standard feature that follows from our explanation of dark
energy as an emergent phenomenon is that the expansion rate in the
early universe is 2.09 times faster than in the models of standard
cosmology (as explained below). This has the consequence that the
deuterium destruction 
process is terminated significantly sooner, leaving a fraction of
undestroyed deuterium that is much higher than the observed abundance,
unless we compensate the faster expansion rate by using a higher value for
the baryon density. Raising $\Omega_B$ increases the deuterium
destruction rate, to allow the same fraction of deuterium to be
destroyed within the shorter time interval that is available for this process. 

Next we will quantify these arguments by showing that an increase of 
the expansion rate by a factor of 2.09 requires an increase of 
$\Omega_B$ to the approximate level of the total matter density
$\Omega_M$ to restore agreement between the BBN 
predictions and the observed deuterium abundance. Since $\Omega_M$
includes all of 
dark matter, we are led to the conclusion that there may be no need to
invoke the existence of non-baryonic matter to explain the observed high value
of $\Omega_M$. All of it can be baryonic without violating the BBN 
constraint imposed by the observed deuterium abundance. 

Here we will limit ourselves to an idealized BBN treatment, since the
full solution of the nuclear rate equations in 
our non-standard cosmology is beyond the scope of the present
paper. Our main idealization is to treat deuterium creation and
destruction as occurring in two separated stages: (1) As soon as
photodissociation vanishes when the expanding universe cools, all of
the free neutrons get captured into 
deuterium nuclei. (2) Subsequently the deuterium destruction begins,
whereby deuterium gets converted into helium via $^3$He or 
tritium. This destruction process occurs with a rate $\gamma$ and 
duration $\Delta t$, after which it ceases, leaving a
surviving deuterium abundance that is a 
factor $\exp(-\,\gamma\,\,\Delta t)$ smaller than the initial value 
at the beginning of stage 2. $\Delta t$ is proportional to the
Hubble time or inverse expansion rate. 

In reality the two stages overlap. The destruction process does not
wait until the creation process is finished, but sets in as soon as a
significant amount of deuterium nuclei have been created. Our
idealization however not only simplifies the calculations, it has the
advantage of bringing out the basic BBN physics in a more transparent way. 

Full BBN calculations show that the surviving deuterium abundance, usually
represented by the number density $y_D
=$ D/H of nuclei relative to hydrogen, depends on the
baryon density $\Omega_B$ in a way that can be approximated in the form
of a power law: $y_D\sim \Omega_B^{-\alpha}$. This interpolation formula gives excellent
results when using $\alpha =1.6$ for values of $\Omega_B$ around 5\,\%,
the value favored in standard cosmology
\citep{stenflo-steigman2007}. However, a glance at Fig.~6.5 in the
monograph by \citet{stenflo-peebles1993} shows that the steepness of
the curve that depicts the calculated deuterium abundance as a function of the
baryon abundance increases significantly with increasing
$\Omega_B$. Near the midpoint (on a logarithmic scale) of the range
between $\Omega_B$ and $\Omega_M$ the slope of the curve 
corresponds to $\alpha\approx 3$. Therefore the interpolation
formula should only be used 
for the purpose of crude estimates if applied to a wider $\Omega_B$
range, as we are doing here. We use it for convenience and because it
makes the presentation
of the physics more transparent. 

With these caveats we will use the interpolation formula to 
treat $\alpha$ as a parameter that gets determined by the assumed
requirement that 
all matter, including dark matter, is baryonic. The consistency of 
this assumption is then tested by checking if the derived value of the
$\alpha$ parameter falls within a physically plausible range around
the representative midrange slope value of $\alpha\approx 3$.  

Let us with the help of this idealization compare the BBN results of standard
and non-standard cosmology, distinguishing them with indices $s$ and $ns$,
respectively. If the two versions of cosmology are both going to agree
with the same observed value for $y_D$, then the following relation
must hold (within the framework of our idealization): 
\begin{equation}
\Omega_{B,s}^{-\alpha}\,\,e^{-\,\gamma\,\Delta t_s}\,=\,\Omega_{B,ns}^{-\alpha}\,\,e^{-\,\gamma\,\Delta t_{ns}}
\,.\label{eq:ydsns}\end{equation}
It describes how any change in the time scale $\Delta t$ must be
compensated for by a corresponding change in the baryon density in
order to preserve agreement with the observations. 

Now let us test if our non-standard cosmology is consistent with a
scenario where all dark matter is baryonic, which means that we set
$\Omega_{B,ns} =\Omega_M$, where $\Omega_M$ represents all matter,
baryonic plus non-baryonic. Removing index $s$ from $\Omega_B$ we 
then obtain from Eq.~(\ref{eq:ydsns}) 
\begin{equation}
\left({\Omega_M\over \Omega_B}\!\right)^{\alpha}\,=\,R_s^{\,\kappa}\,,
\label{eq:ommoveromb}\end{equation}
where 
\begin{equation}
R_s\equiv e^{-\,\gamma\,\Delta t_s}\,,
\label{eq:rs}\end{equation}
and 
\begin{equation}
\kappa \,\equiv\, {\Delta t_{ns}\over \Delta t_s}\,-\,1\,.
\label{eq:kappa}\end{equation}
According to \citet{stenflo-planck2016} $\Omega_M/ \Omega_B =6.3$. 
The destruction factor $R_s$ is the ratio between the final value of
$y_D$ and its initial value $y_D^{(0)}$ at the beginning of stage 2
in our idealized scenario. If all free neutrons at the start of stage
1 end up as part of $^4$He with a mass fraction $Y$ and do it
after first having been absorbed into deuterium, then the mass
fraction of deuterium at  the beginning of stage 2 is also
$Y$. Expressing $Y$ in terms of the number density $y_D^{(0)}$
relative to hydrogen, we have $Y=2 y_D^{(0)}/(2y_D^{(0)}+1)$, which,
when inverted, gives 
\begin{equation}
y_D^{(0)} = {0.5Y\over 1-Y}\,.
\label{eq:yd0}\end{equation}
According to \citet{stenflo-steigman2007} $y_D = 2.6\times 10^{-5}$
and $Y=0.249$, which according to Eq.~(\ref{eq:yd0}) gives
$y_D^{(0)}=0.166$. We then find $R_s =y_D/y_D^{(0)}=1.57\times
10^{-4}$, which demonstrates that only a minor fraction of the
deuterium survives destruction in the Big Bang. 

The ratio $\Delta t_{ns}/ \Delta t_s$ equals the ratio between the
corresponding Hubble times. At the end of Sect.~\ref{sec:linkplanck}
we showed that in the radiation-dominated era of our non-standard
cosmology the Hubble time is 95.8\,\%\ of the age of the universe,
while in standard cosmology it is 2.0 times the age. The ratio $\Delta
t_{ns}/ \Delta t_s$ is therefore $0.958/2.0=0.479$, which gives us
$\kappa =-0.521$. 

As we have now assigned observationally constrained values for all the parameters in
Eq.~(\ref{eq:ommoveromb}) except for the parameter
$\alpha$, we can solve for $\alpha$, obtaining 
\begin{equation}
\alpha =2.5\,.
\label{eq:betasolexp}\end{equation}
This value should be compared with the previously mentioned midrange
slope value of $\alpha\approx 3$ that is representative for the
actual slope derived from rigorous BBN calculations. In view of the
uncertainties of our simplified treatment the agreement
between the two values is sufficiently close to satisfy our consistency test:
With the $1/0.479=2.09$ times faster expansion rate of our theory, agreement
between the BBN calculations and the observed deuterium abundance gets
restored if the baryon density parameter $\Omega_B$ is raised to the
level of the total matter density parameter $\Omega_M$. 
This suggests that all dark matter may indeed be
baryonic without violating BBN,
there may be no reason to introduce yet to be discovered exotic particles
for the purpose of explaning the observed deuterium abundance. Because 
of our simplified treatment, however, this conclusion still needs to be
validated by full and rigorous BBN calculations. 

If all dark matter is indeed baryonic, it is clear that most of it
must be cold and collisionless, and therefore have macroscopic
properties that are not that different from those of WIMPs. Such
behavior would be the case if it for instance would be composed of grains, 
rocks, and primordial black holes, with a spectrum of sizes spanning from tiny grains to
planetary size bodies. Larger dark bodies (MACHO ---
Massive Astrophysical Halo Objects) appear to be
disfavored by constraints from gravitational microlensing as major
candidates for dark matter, although the evidence is not yet conclusive 
\citep[cf.][]{stenflo-novati2005,stenflo-tisserand2007}. A high value of $\Omega_B$ 
poses other issues beyond  
BBN, which need to be clarified, for example compatibility with the
observed signatures of baryon acoustic oscillations (BAO) in the CMB
spectrum. The CMB imprint is in the form of a characteristic distance
scale that represents the sound horizon, the comoving distance that sound
waves can travel from the time of the Big Bang to the time when the
baryons decouple from the radiation field. Like in the BBN case we
have two competing effects. The speed of sound depends on the
baryon density, while the travel time depends on the expansion rate of
the universe. Both are modified in our non-standard cosmology. It is
however outside the scope of the present paper to work out the details
of this here.

\section{Conclusions}\label{sec:concl}
Our resolution of both the dark energy and dark matter conundrums has
been achieved without the use of any free parameters and without any
modification of the Einstein equations for gravity, except for the
removal of the ``cosmological constant''  from these equations. The
origin of the observed accelerated expansion of the universe is not in
the formulation of the equations for gravity, but has to do with the cosmic
boundary conditions. 

The cosmological phenomena that are generally referred to
with the label ``dark energy'' may be seen as the result of a global
cosmic resonance that emerges because the age of the universe is
finite. The $\Lambda$ 
term has the dimension of the square of 
a frequency, which we may think of as the ``pitch'' of the
universe. In the beginning, when the horizon of the universe
was small, the pitch was high, but it got lower as the universe
increased in size, in inverse proportion to the horizon radius. 

In contrast, when $\Lambda$ is put in by hand as a cosmological
constant, the pitch always remains the same, regardless of whether the
universe is small or large. Since there is no physical justification
for inserting such a constant, the anthropic principle has often been 
invoked in the guise of an ``explanation'': The existence of
biological life constrains the allowed values of 
$\Lambda$ to a narrow range around the actually observed value. Such
an argument 
opens the door to the proliferation of parallel universes with different
values of the cosmological constant, some of which are harbouring
life, while most of them do not. 

According to the present work the possibility of
universes with other values of $\Lambda$ does not exist. Nature did
not have a choice, because the requirement of logical consistency
leads to uniqueness. 

The explanation of dark energy as an emergent phenomenon leads to
non-standard cosmological consequences. One of
these consequences provides a resolution of the dark matter
enigma. This resolution is not in terms of
modified gravity, because this would require a major modification at
scales several orders of magnitude smaller than the horizon scale, for
which there is no justification. Instead 
dark matter must really be made up of physical particles. However,
because of the non-standard cosmology that follows from our
explanation of dark energy, all the dark matter 
particles now may be baryonic, there may not be any need to invoke the
existence of some yet to be discovered exotic particles (WIMPs). 

The main reason for the belief in the existence of non-baryonic matter
has come from the comparison of the observed abundances of the light
elements with the predictions from BBN (Big Bang nucleosynthesis)
calculations. While the relative fraction $\Omega_M$ of matter in the
universe is of order 30\,\%, at most 5\,\%\ can be baryonic ($\Omega_B$) to satisfy
the BBN constraints. 

These constraints are however based on the framework of standard
cosmology. The non-standard aspects in our theory for dark energy lead
to an expansion rate in the early universe that is 2.1 times the
expansion rate in the standard cosmological models used for the
BBN calculations. When the faster value of the expansion rate is used
for the calculations, the baryonic mass fraction $\Omega_B$ must be
increased to the level around that of $\Omega_M$ to be compatible with
the observed abundances. To show this we have used a simplified treatment
focused on the case of the deuterium abundance, so this resolution
of the dark matter enigma still needs to be validated by more complete
and rigorous modeling of BBN and other relevant observational
constraints, like the observed CMB imprints of the baryon acoustic oscillations. 

Our explanation of dark energy uses classical theory, at least in the
sense that Planck's constant does not appear in the expression for
$\Omega_\Lambda$ in 
Eq.~(\ref{eq:omlfinal}). Our conclusion that dark matter is baryonic does
not make use of anything beyond the well-established domain of
particle physics, no ``exotic physics'' is called for. Nevertheless
the process of clarifying the role of the cosmic boundary conditions 
has cast some light on intriguing aspects of gravity. Examples: the wave nature of
the gravitational interaction, the feedback effects of the vacuum
energy (vacuum polarization), structural similarities with quantum
field theory, metric-induced thermodynamics, scaling
relations between the present and the Planck era, and the necessity of
a hot Big Bang from Heisenberg's uncertainty principle. These aspects
may help guide us in our quest for a theory of quantum gravity.

\begin{acknowledgements}
I am grateful to Philippe Jetzer for helpful comments on the
  manuscript. 
\end{acknowledgements}


\end{document}